\documentclass[11pt]{article}
\pdfoutput=1
\usepackage[pdftex,bookmarks,colorlinks]{hyperref}
\begin{document}
\title{Whipping Instabilities in Electrified Liquid Jets}
\author{\'Alvaro G. Mar\'in$^1$\footnote{ Corresponding Author: a.g.marin@utwente.nl},
Guillaume Riboux $^1$, \\Ignacio G. Loscertales $^2$, Antonio
Barrero $^1$\\\,
\\\vspace{6pt}$^1$Department of Aerospace Engineering, University of Seville
\\\vspace{6pt}$^2$Department of Mechanical Engineering, University of
M\'alaga}
\maketitle

\begin{abstract}
A liquid jet may develop different types of instabilities, like
the so-called \textit{Rayleigh-Plateau instability}, which breaks
the jet into droplets. However, another types of instabilities may
appear when we electrify a liquid jet and induce some charge at
his surface. Among them, the most common is the so-called
\textit{Whipping instability}, which is characterized by violent
and fast lashes of the jet. In the submitted fluid dynamic video
\href{http://hdl.handle.net/1813/11422}{Video}\cite{movideos} we
will show an unstable charged glycerine jet in a dielectric liquid
bath, which permits an enhanced visualization of the instability.
For this reason, it is probably the first time that these
phenomena are visualized with enough clarity to analyze features
as the effect of the feeding liquid flow rate through the jet or
as the surprising spontaneous stabilization at some critical
distance to the ground electrode.

\end{abstract}


\section{Introduction}

A charged liquid jet may develop several types of instability, but
among them, the so-called \emph{Whipping instability}. This
instability is the base of the popular technique called
\emph{Electrospinning} employed for the generation of polymer
nanofibers. The instability analyzed by G. I. Taylor
\cite{taylorjets} among others, who unfortunately gave a wrong
explanation to the instability: \emph{"Their stability }(liquid
charged jets) \emph{seems to be due to a mechanical rather than
electrical causes, like that of a stretched string, which is
straight when pulled but bent when pushed"}. Taylor concluded that
air friction acted as an "pulling" force opposing that
"stretching" effect of the electrical force applied in the
streamwise direction. And he continued in the same article:
\emph{"I must confess that the experiments to be described fail
also, but} [...] \emph{it seems worth publishing these photographs
with the relevant data in the hope that someone may give a
relevant analytical description of the }[...] \emph{unstabilizing
effect on disturbances which displace the jet laterally"}.
Although the effect pointed out by Taylor of air/liquid friction
is probably not negligible, nowadays it is well established that
the origin of the instability is more related to the
self-repulsion of the jet due to its high surface charge. The
model developed by Reneker et al. \cite{reneker} illustrates this
effect: If a line of electrical charges of the same sign flows in
the same direction they will tend to separate from each other in
order to reduce the energy of the system. If there is additionally
a force binding them altogether, then it can be demonstrated that
the most energetically favorable path to follow is an helicoidal
path, as real charged jets certainly do. This is the basic idea of
the model developed by Reneker et al. whose simulations
successfully describe the general aspects of the instability.
However, due to his treatment of the jet as a discrete line of
charges, it is extremely complicated to make a connection with
charged liquid jets in the experiments. Despite of the amount of
literature on the topic, it is not still clear which are the
critical parameters triggering the instability. The difficulty in
the experimental research relies in the violence of the
instability, usually manifested with fast lashes at large scales.
However, we have found that when the experiment is performed
inside a dielectric liquid medium, the instability develops in a
more reduced scale and follows some ordered patterns. This
advantage permits us to visualize and film the instability with a
resolution as it has never been observed before.

\section{The experiments}
We proceed to describe the video included in the reference
\href{http://hdl.handle.net/1813/11422}{Video}\cite{movideos}. In
the first three filmed experiments we show the effect of the
injected flow rate through an electrically driven jet of
glycerine. As we increase the amount of supplied liquid, the jet
not only increases in diameter and in total length, but also the
total net charge transported by its surface is enhanced. In
consequence, the jet becomes more unstable manifesting higher
frequencies and non-periodic motion in the extremes of the jet, as
we can see in the third experiment.
\\

Another interesting aspect is the ability of the jets to turn
suddenly stable at certain distances from the electrode, which is
presented in the second part of the video. Some mechanisms have
been proposed for this surprising stability based on the
electrical discharge of the jet close to the electrodes
\cite{saville}. Unfortunately, the proposed explanations can not
be applicable in this case since our experiments have been all
performed inside a dielectric liquid medium (hexane), therefore we
can assert that there must be a different mechanism behind this
phenomenon. Finally, at the end of the video we show how the
electrified micrometric jet surprisingly coils in the liquid
electrode surface, just as uncharged liquid viscous filaments do
after impacting upon a surface. Most of the relevant technical
details are described in the videos, and a detailed analysis of
all these experiments will be presented by Riboux et al. in the
APS-DFD Annual Meeting 2008 \cite{guille}.

\end{document}